\documentclass[a4paper,11pt]{article} 
\usepackage[margin=0.5in]{geometry} 
\usepackage{amssymb}
\usepackage{amsmath,epsfig, epsf, epic, graphicx} 
\usepackage{fancybox, lscape, subfigure} 
\usepackage{bm}
\usepackage{natbib}
\usepackage{float}
\usepackage[margin=2cm]{caption}
\usepackage{setspace}
\usepackage[usenames,dvipsnames]{color}

\title{High-dimensional confounding adjustment using continuous spike and slab priors}
\author{Joseph Antonelli, Giovanni Parmigiani, Francesca Dominici}

\begin{document}
\date{}
\maketitle{}

\abstract{In observational studies, estimation of a causal effect of a treatment on an outcome relies on proper adjustment for confounding. If the number of the potential confounders ($p$) is larger than the number of observations ($n$), then direct control for all potential confounders is infeasible. Existing approaches for dimension reduction and penalization are generally aimed at predicting the outcome, and are less suited for estimation of causal effects. Under standard penalization approaches (e.g. Lasso), if a variable $X_j$ is strongly associated with the treatment $T$ but weakly with the outcome $Y$, the coefficient $\beta_j$ will be shrunk towards zero thus leading to confounding bias.
Under the assumption of a linear model for the outcome and sparsity, we propose continuous spike and slab priors on the regression coefficients $\beta_j$ corresponding to the potential confounders $X_j$.  Specifically, we introduce a prior distribution that does not heavily shrink to zero the coefficients ($\beta_j$s) of the $X_j$s that are strongly associated with $T$ but weakly associated with $Y$. We compare our proposed approach to several state of the art methods proposed in the literature. Our proposed approach has the following features: 1) it reduces confounding bias in high dimensional settings; 2) it shrinks towards zero coefficients of instrumental variables; and 3) it achieves good coverages even in small sample sizes. We apply our approach to the National Health and Nutrition Examination Survey (NHANES) data to estimate the causal effects of persistent pesticide exposure on triglyceride levels.

{High-dimensional data; Causal inference; Bayesian variable selection; Shrinkage priors}}

\section{Introduction}
In observational studies, we are often interested in estimating the causal effect of a treatment $T$ on an outcome $Y$, which requires proper adjustment of a set of potential confounders $\boldsymbol{X}$. In the context of high-dimensional data, where the number of potential measured confounders $p$ could be even larger than the sample size $n$, standard methods for confounding adjustment such as regression or propensity scores \citep{rosenbaum1983central} will fail.

In the context of prediction, a variety of methods exist for imposing sparsity in regression models with a high-dimensional set of covariates. Arguably the most popular, the lasso \citep{tibshirani1996regression} places a penalty on the absolute value of the coefficients from a regression model, thus shrinking many of them to be exactly zero, leading to a more parsimonious model. A variety of extensions to the lasso have been proposed such as the SCAD, elastic net, and adaptive lasso penalties, to name a few \citep{fan2001variable, zou2005regularization,zou2006adaptive}. One challenge encountered with all of these approaches is the difficulty to provide a meaningful assessment of uncertainty around estimates of the regression coefficients. While progress has been made recently on this topic \citep{lockhart2014significance,taylor2016post}, it remains difficult to obtain valid confidence intervals for parameters under complex, high-dimensional models. 

Bayesian models can alleviate these issues by providing valid inference from posterior samples. Much of the recent work has centered around shrinkage priors, which can be represented as scale mixtures of Gaussian distributions and allow for straightforward posterior sampling. \cite{park2008bayesian} introduced the Bayesian lasso: a scale mixture of Gaussians with an exponential mixing distribution that induces wider tails than a standard normal prior. More recently, global-local shrinkage priors have been advocated that have a global shrinking parameter that applies to all parameters, as well as local shrinking parameters which are unique to the individual coefficients. \cite{carvalho2010horseshoe} introduced the horseshoe prior, which is a scaled mixture of Gaussians with a half-cauchy mixing distribution that has been shown empirically to have good performance in high-dimensional settings. \cite{bhattacharya2015dirichlet} introduced a new class of distributions that are also scaled mixtures of Gaussians with an additional Dirichlet mixing component and proved that it's posterior concentrates at the optimal rate. \cite{rovckova2016spike} differ somewhat in that they adopt the spike and slab formulation of \cite{george1993variable}, however both the spike and slab priors are Laplace distributions. All of these approaches are aimed at obtaining ideal amounts of shrinkage in high-dimensional settings where large coefficients should be shrunken a small amount, while others are shrunken heavily towards zero.

These and many other approaches have been based on the same principles of aiming to reduce shrinkage for important covariates in the context of prediction of $Y$. Several authors have pointed out that frequentist and Bayesian procedures for variable selection or shrinkage that focus on predicting $Y$ perform poorly when the inferential goal is estimation of the effect of $T$ on $Y$ \citep{crainiceanu2008adjustment, wang2012bayesian, belloni2013inference, belloni2017program,hahn2017bayesian}.A variety of data driven methods have been developed to select confounders in causal inference \citep{van2010collaborative,de2011covariate,vansteelandt2012model,wang2012bayesian,zigler2014uncertainty}. Many of these approaches rely on the specification of a treatment model $E(T \vert \boldsymbol{X})$, and an outcome model $E(Y \vert T, \boldsymbol{X})$. \cite{wang2012bayesian} introduced a Bayesian model averaging approach for estimating the effect of $T$ on $Y$ averaged across models that include different sets of potential confounders. They assume a priori that if a covariate $X_j$ is associated with the treatment $T$ then this covariate should have high probability to be included into the outcome model, even if this covariate is weakly associated with the outcome. Many ideas have been built on this prior specification to address the issue of confounder selection and model uncertainty \citep{talbot2015bayesian, wang2015accounting,cefalu2016model,antonelli2017gbac}. All of the aforementioned approaches have been shown to work well in identifying confounders or adjusting for confounding; however, none of these approaches are well-suited to a high-dimensional vector of confounders. Recently, there has been increased attention to estimating treatment effects when $p \geq n$. \cite{wilson2014confounder} introduced a decision theoretic approach to confounder selection for $p \geq n$. They showed that their approach has connections to the adaptive lasso, but with weights aimed at reducing shrinkage of confounders, rather than predictors. \cite{belloni2013inference} applied standard lasso models on both the treatment model $E(T \vert \boldsymbol{X})$, and the outcome model $E(Y \vert T, \boldsymbol{X})$, separately. Then, they identify as confounders the union of the variables that were not shrunk to zero in the two models, and estimate the causal effect using this reduced set of covariates. \cite{farrell2015robust} first applies lasso models to treatment and outcome models to select confounders. Second, they calculate a double robust estimator using the resulting unpenalized treatment and outcome models. \cite{antonelli2016double} implemented a similar doubly robust estimation approach using standard lasso outcome and treatment models but in context of matching on both the propensity and prognostic scores. \cite{ertefaie2013variable} proposed an alternative approach for selecting confounders in high-dimensional settings by penalizing a joint likelihood for both the treatment and the outcome model to ultimately lead to the selection of important confounders. \cite{shortreed2017outcome} used similar ideas by fitting an adaptive lasso to a propensity score model, and show that it leads to the inclusion of only covariates necessary for confounding adjustment or outcome model prediction.   \cite{hahn2016bayesian} utilized horseshoe priors on a re-parameterized likelihood that aims to reduce shrinkage for important confounders. \cite{athey2016approximate} combined high-dimensional regression with the balancing weights of \cite{zubizarreta2015stable} to obtain valid inference of treatment effects even when the true data generating models are not sparse. All of these approaches have the advantage of being able to handle settings with  $p \geq n$, however, as we will demonstrate in simulations, existing approaches relying on asymptotic theory can provide coverage below the nominal level in finite samples. Also, many of these approaches will tend to include instrumental variables, are not applicable to studying the effects of continuous treatments (see Section \ref{sec:nhanes}), or both. 


In Section \ref{sec:model} we propose spike and slab priors for confounding adjustment in the context of homogeneous and heterogeneous treatment effects. In section \ref{sec:computation}, we detail the Bayesian computations, including the selection of tuning parameters to achieve a good compromise between sparsity and addressing confounding bias. In section \ref{sec:sims}, we present results from several simulation studies that include homogeneous and heterogeneous treatment effects, strong and weak confounders, instrumental variables, and sparse and non sparse settings. In Section \ref{sec:nhanes}, we present the data analysis which considers continuous treatments, In section \ref{sec:discussion} we conclude with a summary of the strengths and weaknesses of the proposed approach and future research directions.


\section{Spike and slab priors for confounding adjustment}
\label{sec:model}

Throughout, we will assume that we observe $\boldsymbol{D_i} = (Y_i, T_i, \boldsymbol{X_i}$) for $i,\dots,n$, where $n$ is the sample size of the observed data, $Y_i$ is the outcome, $T_i$ is the treatment, and $\boldsymbol{X_i}$ is a p-dimensional vector of pre-treatment covariates for subject $i$. We will assume for simplicity that $Y_i$ is continuous, though we will not make assumptions regarding $T_i$, as it can be binary, continuous, or categorical. The extension to binary outcomes is straightforward using latent variable techniques introduced in \cite{albert1993bayesian}. In general we will be working under the high-dimensional scenario of $p \geq n$, where we let $p \rightarrow \infty$. Our estimand of interest is the average treatment effect (ATE), defined as $\Delta(t_1, t_2) = E(Y(t_1) - Y(t_2))$, where $Y_i(t)$ is the potential outcome subject $i$ would receive under treatment $t$. We will assume that the probability of receiving any value of treatment is greater than 0 for any combination of the covariates, commonly referred to as positivity. We make the stable unit treatment value assumption (SUTVA) \citep{little2000causal}, which states that the treatment received by one observation or unit does not affect the outcomes of other units and the potential outcomes are well-defined. We will further assume strong ignorability conditional on the observed covariates, and that the covariates necessary for ignorability are an unknown subset of $\boldsymbol{X}$. Strong ignorability implies that potential outcomes are independent of $T$ conditional on $\boldsymbol{X}$. 

\subsection{Model formulation}

In this section we assume a homogeneous treatment effect, i.e that the treatment effect is the same across all values of the covariates $\boldsymbol{X}$. We will relax this assumption in Section \ref{sec:hetero}. We introduce the following hierarchical formulation: 
\begin{align}
	Y_i\mid T_i, \boldsymbol{X_i}, \beta_0, \beta_t, \boldsymbol{\beta}, \sigma^2 &\sim \text{Normal}(\beta_0 + \beta_t T_i + \boldsymbol{X}_i\boldsymbol{\beta}, \sigma^2) \label{eqn:outcome} \\
	P(\boldsymbol{\beta} \vert \boldsymbol{\gamma}) &= \prod_{j=1}^p \gamma_j \psi_1(\beta_j) + (1 - \gamma_j) \psi_0(\beta_j) \nonumber \\
	P(\boldsymbol{\gamma} \vert \theta) &= \prod_{j=1}^p \theta^{w_j \gamma_j} (1 - \theta^{w_j})^{1 - \gamma_j} \nonumber \\
	P(\theta \vert a,b) &\sim \text{Beta}(a,b) \nonumber \\
	P(\sigma^2 \vert c,d) &\sim \text{InvGamma}(c, d) \nonumber \\
	P(\beta_0), P(\beta_t) &\sim \text{Normal}(0, K). \nonumber 
\end{align}

\noindent Under these assumptions, $\Delta(t_1, t_2) = (t_1 - t_2) \beta_t$. It is straightforward to allow the treatment effect to be nonlinear by replacing $\beta_t T_i$ with $f(T_i)$, which can be approximated using basis functions. If $\gamma_j=1$ then $\beta_j \sim  \psi_1(\beta_j)$ - the slab component of the prior. If $\gamma_j=0$ then  
$\beta_j \sim  \psi_0(\beta_j)$ -- the spike component of the prior.  Therefore $\gamma_j=1$ indicates that  $X_j$ is potentially an important confounder. We set $\psi_1(\cdot)$ and $\psi_0(\cdot)$ to be Laplace distributions with densities $\psi_1(\beta_j) = \frac{\lambda_1}{2 \sigma}e^{-\frac{\lambda_1 |\beta_j|}{\sigma}}$ and $\psi_0(\beta_j) = \frac{\lambda_0}{2 \sigma}e^{-\frac{\lambda_0 |\beta_j|}{\sigma}}$, respectively. More specifically, when $\gamma_j=1$, the prior standard deviation of $\beta_j$ is $\sigma \sqrt{2}/\lambda_1$ and when $\gamma_j = 0$ the prior standard deviation is $\sigma \sqrt{2}/\lambda_0$. Scaling the prior variance with $\sigma^2$ is common in Bayesian hierarchical models \citep{park2008bayesian} and allows for more stable estimation and better interpretation of $\lambda_1$ and $\lambda_0$. Finally, we have $\theta$ and $w_j$, which control the prior probability that $\gamma_j = 1$. The global parameter $\theta$ dictates the probability that $\gamma_j = 1$ when $w_j = 1$ and can be thought of as the overall sparsity level in the data. The weights $w_j$ are tuning parameters that we will use to prioritize variables to have $\gamma_j = 1$ if they are also associated with the treatment. We will discuss the selection of $w_j$ in more detail in Sections \ref{sec:probslab} - \ref{sec:delta}.

\subsection{Hyper prior selection}

Our prior formulation has a number of hyper parameters, and it is important for these to be set to reasonable values to obtain good inference for the treatment effect of interest. The first hyper parameters are $\lambda_0$ and $\lambda_1$, which control the variance of the spike and slab priors. Following \cite{rovckova2016spike}, we will fix $\lambda_1$ to a small value, say 0.1, so that the prior variance for coefficients in the slab component of the prior is high enough to be reasonably uninformative, and important parameters will not be shrunk heavily towards 0. We assess the sensitivity to this choice in the supplementary materials and find that results are robust to the choice of $\lambda_1$. Results can be quite sensitive to the choice of $\lambda_0$, therefore we will estimate it using empirical Bayes to let the data determine how much to shrink coefficients that are placed in the spike component of the prior.

The parameters $a$ and $b$ dictate the prior for $\theta$ meaning they control the amount of sparsity induced a priori. We will adopt standard practice in the high-dimensional Bayesian literature and set $a$ to a constant, and set $b \propto p$ \citep{Zhou2014,rovckova2016spike}. This prior more aggressively shrinks coefficients to the spike component of the prior as $p$ grows. This feature is desirable in high-dimensional models where we must more aggressively shrink parameters as the covariate space grows to avoid the curse of dimensionality \citep{scott2010bayes}. Throughout the paper we will use $a=1$ and $b=0.1p$, though we assessed the sensitivity to these choices in the supplementary materials and found the results are robust to the selection of $a$ and $b$. We will assume conjugate and uninformative priors for  $\sigma^2$, $\beta_0$ and $\beta_t$. Finally, we must choose the tuning parameter, $w_j$, which we will use to prioritize potential confounders in the prior formulation. We will discuss the selection of $w_j$ in Section \ref{sec:delta}.

\subsection{Probability of inclusion into the slab}
\label{sec:probslab}

To better understand how to select $w_j$, it helps to understand the probability that a parameter for a given covariate is included in the slab component of the prior. There are two crucial quantities that we can study to gain intuition into whether an important covariate is effectively included in the model. The first is the conditional probability that a parameter is included in the slab component of the prior, which can be defined as follows:

\begin{align}
	p_{\theta}^*(\beta_j) = P(\gamma_j = 1 \vert \beta_j, \theta) = \frac{\theta^{w_j} \psi_1(\beta_j)}{\theta^{w_j} \psi_1(\beta_j) + (1 - \theta^{w_j}) \psi_0(\beta_j)}. \nonumber
\end{align}

This is also the expression from which we update $\gamma_j$ in a gibbs sampler, and therefore gives insight into the probability that a parameter is included in the slab component of the prior. The second quantity involves the posterior mode of our model. As seen in \cite{rovckova2016spike} the posterior mode will be sparse in the sense that many of the parameters will be set exactly to zero. An important quantity in the estimation of the posterior mode is defined as

\begin{align}
	\Delta_j \equiv \underset{t>0}{inf} \left\{ nt/2 + \frac{\lambda_1 t - \log \left( \frac{p_{\theta}^*(0)}{p_{\theta}^*(t)} \right)}{t} \right\}. \nonumber
\end{align}

\begin{figure}[h]
\centering
 \includegraphics[width=0.9\linewidth]{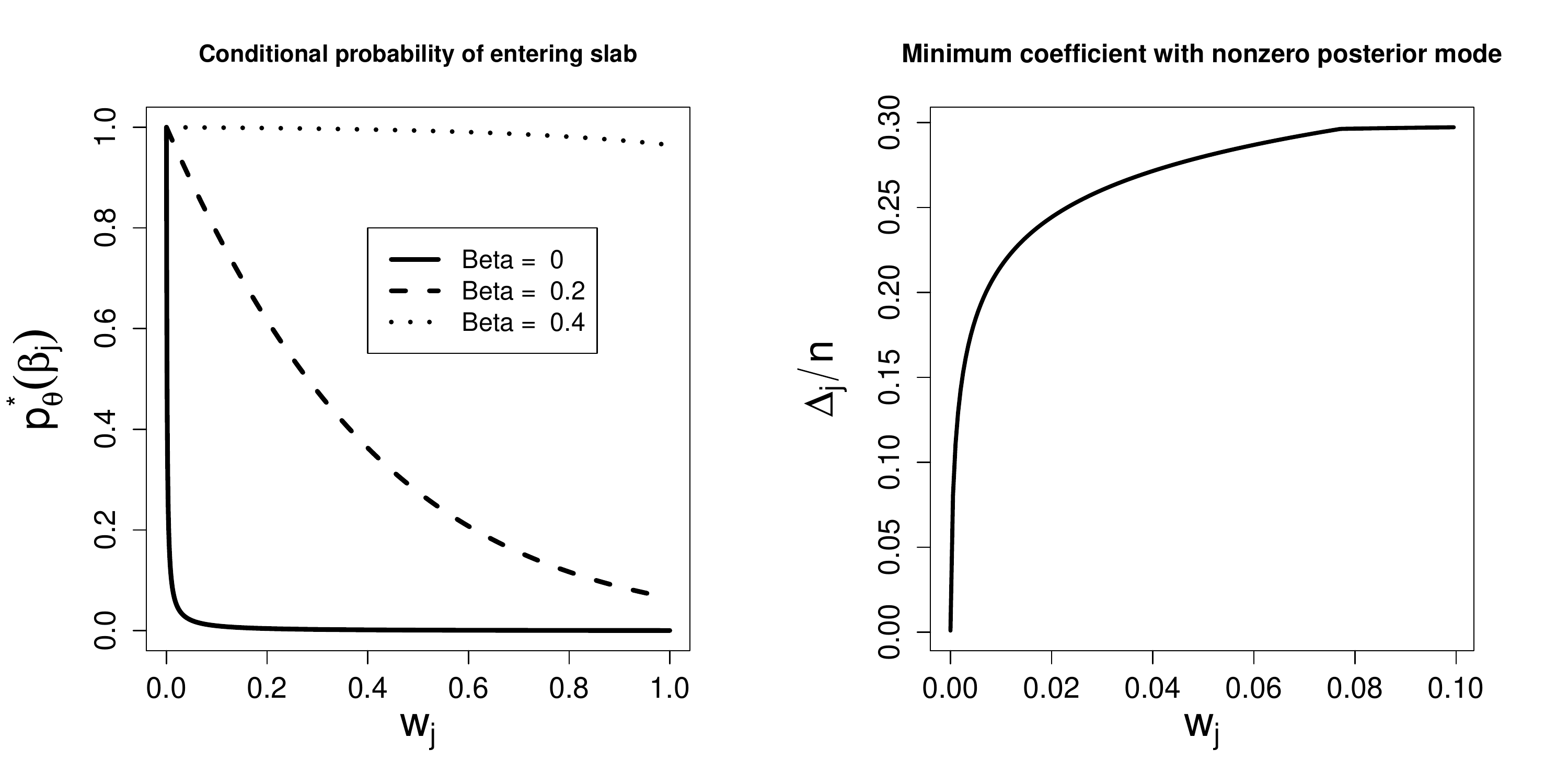}
\caption{The left panel shows $p_{\theta}^*(\beta_j)$ for a variety of values of $\beta_j$ as a function of $w_j$. The right panel shows $\Delta_j / n$ as a function of $w_j$. Here we fixed $\lambda_1 = 0.1$, $\lambda_0=30$, and $\theta=0.05$.}
\label{fig:PaperProbs}
\end{figure}

This expression is important because the posterior mode estimate of $\widehat{\beta}_j$ is nonzero if $\boldsymbol{X_j}'(\boldsymbol{Y} - \sum_{l \neq j} \boldsymbol{X_l}\widehat{\beta_l} - \boldsymbol{T} \widehat{\beta_t} - \widehat{\beta}_0) > \Delta_j$. Now that we have defined these two quantities, we can look at them with respect to $w_j$ to gain some intuition as to how $w_j$ impacts the variable selection and subsequent shrinkage of important parameters. Figure \ref{fig:PaperProbs} shows these two quantities as a function of $w_j$ when $\lambda_1 = 0.1$, $\lambda_0=30$, and $\theta=0.05$. The left panel shows that covariates strongly associated with the outcome ($\beta_j = 0.4$) always enter the slab regardless of $w_j$, while variables with no association ($\beta_j = 0$) never enter the slab unless $w_j$ is very close to 0. Covariates with a mild association with the outcome ($\beta_j = 0.2$) change drastically depending on $w_j$, as small values of $w_j$ lead to inclusion probabilities near 1 and large values of $w_j$ lead to posterior inclusion probabilities near 0. The right panel of Figure \ref{fig:PaperProbs} shows that the threshold for a parameter having a nonzero posterior mode greatly decreases when $w_j$ is small. Not seen in the figure is that values of $w_j$ greater than 0.1 are essentially the same as $w_j = 1$ in terms of the probability of being nonzero in the posterior mode.

\subsection{Selection of $w_j$}
\label{sec:delta}
In the previous section we saw how the weight $w_j$ can impact variable selection as it is decreased towards 0. Our goal is improved estimation of $\beta_t$, the treatment effect, and therefore we want to prioritize variables that are also associated with the treatment, $T_i$. If a variable $X_j$ is associated with $T$, then omitting $X_j$ in the outcome model could lead to confounding bias. Consequently, it is desirable to increase the prior probability that $\beta_j$ is in the slab component of the prior. With this guiding principle in mind, we do the following: 1) we use lasso to fit the exposure model $E(T \vert \boldsymbol{X})$ \citep{tibshirani1996regression}; 2)
for each $X_j$ that has a non zero regression coefficient from the lasso estimation of the exposure model, we set $w_j=\delta$  where $0 < \delta < 1$. Please note that if $\delta<1$, then  $\theta^{\delta} > \theta$ which leads to a higher prior probability for $\beta_j$ to be included into the slab component of the prior. Smaller values of $\delta$ lead to more protection against omitting an important confounder. However, values of $\delta$ too small might lead to inclusion of instrumental variables which decrease efficiency and can amplify bias in the presence of unmeasured confounding \citep{pearl2011invited}. 

Now we provide guidance on how to select a reasonable value of $\delta$. Figure \ref{fig:PaperProbs} can be used to guide our choice of $\delta$. We assume $w_j=\delta\; \forall j$, and we want to set $\delta$ to be as small as possible to protect us against shrinking the coefficients for important confounders, but we also want to ensure that coefficients for instrumental variables or noise variables are heavily shrunk towards 0. We can see in Figure \ref{fig:PaperProbs} that we can set $\delta$ to be as small as possible to increase $p_{\theta}^*(\beta_j)$ for variables with moderate associations with the outcome, while still keeping $p_{\theta}^*(0)$ low. One possibility is to select the minimum value of $\delta$ such that $p_{\theta}^*(0)$ is less than some threshold, such as 0.1. This would imply that the probability of including a parameter for an instrument or noise variable into the slab component of the prior would be 0.1. Intuitively, this threshold represents the point at which we can get the most protection against residual confounding bias while alleviating the impact of instrumental variables. 

Additionally, we have found that when the treatment assignment is not sparse, assigning weights of $\delta$ to all covariates identified by a treatment model can lead to poor performance in the subsequent outcome model. One approach to these types of issues is to cap the number of variables that are prioritized by the treatment model to $k$ covariates. We will explore a scenario where this is the case in the simulation study of Section \ref{sec:sims} and assess the extent that this problem is corrected when we limit the number of variables prioritized.

\subsection{Heterogeneous treatment effects}
\label{sec:hetero}
In this section we describe our approach under the more general case of treatment effect heterogeneity and in the context where the treatment variable is binary or categorical. Addressing treatment effect heterogeneity in the presence of continuous treatments is a more difficult problem, which is beyond the scope of this paper. In the case of a binary treatment, we now specify the same model as in Section \ref{sec:model} but separately for $t=1$ and $t=0$:


\begin{align}
Y_i\mid T_i = t, \boldsymbol{X_i}, \beta_0^{(t)}, \boldsymbol{\beta}^{(t)}, {\sigma^2}^{(t)} &\sim \text{Normal}(\beta_0^{(t)} + \boldsymbol{X}_i  \boldsymbol{\beta}^{(t)}, {\sigma^2}^{(t)}). \nonumber
\end{align}

It is important to note that for each treatment level $t$, we fit a separate model only using subjects $i$ with $T_i = t$. To estimate the treatment effect in this setting, we can exploit the fact that the average treatment effect can be estimated as

\begin{align}
	\Delta(t_1, t_2) &= \int_{\boldsymbol{X}} E(Y \vert T=t_1, \boldsymbol{X}) - E(Y \vert T=t_2, \boldsymbol{X}) dF_n(\boldsymbol{X}) \nonumber
\end{align}

where $F_n(\boldsymbol{X})$ is the empirical distribution of the covariates. If we let $s$ denote the $s^{\text{th}}$ posterior draw obtained in our MCMC algorithm then the posterior mean of the treatment effect can be defined as

\begin{align}
	E(\Delta(t_1, t_2) \vert \boldsymbol{D}) = \frac{1}{Sn} \sum_{s=1}^S \sum_{i=1}^n \beta_{0 (s)}^{(t_1)} - \beta_{0 (s)}^{(t_2)} + \boldsymbol{X_i} (\boldsymbol{\beta}_{(s)}^{(t_1)} -  \boldsymbol{\beta}_{(s)}^{(t_2)}). \nonumber
\end{align}

This provides us with a valid estimate of the treatment effect, however, it does not provide us with a valid credible interval. This estimate is marginalizing over the covariates, and our posterior distribution does not take into account this additional uncertainty, so we will utilize the Bayesian bootstrap \citep{rubin1981bayesian} to account for it. Specifically, we can define $u_0 = 0$, $u_n = 1$, and $u_1$ through $u_{n-1}$ to be the order statistics from $n-1$ draws from a uniform distribution. Then we can define weights, $\xi_i = u_i - u_{i-1}$. We will do this $M$ separate times leading to weights $\xi_{mi}$ for $m=1,\dots, M$ and $i = 1, \dots, n$. Finally, for each of the $S$ posterior samples and $M$ weight vectors we can calculate

\begin{align}
	\Delta_{(s,m)}(t_1, t_2) = \frac{1}{n} \sum_{i=1}^n \xi_{mi} \left( \beta_{0 (s)}^{(t_1)} - \beta_{0 (s)}^{(t_2)} + \boldsymbol{X_i} (\boldsymbol{\beta}_{(s)}^{(t_1)} -  \boldsymbol{\beta}_{(s)}^{(t_2)}) \right) \label{eqn:hetero}.
\end{align}

and we can use the quantiles of these values to create credible intervals. In brief, we have built separate regression models for each of the treatment levels and then taken the difference in the mean predicted values from these models for each observation in the data. To account for the additional uncertainty from marginalizing over covariates, we randomly re-weighted the data using the Bayesian bootstrap. If we were interested in estimating treatment effects within particular subgroups such as the treatment effect on the treated, then the sum in equation \ref{eqn:hetero} would be over only those subjects of interest. Note that while we have separated the estimation of the models for each treatment group, it is possible to posit a hierarchical model to borrow information between treatment groups. This could exploit the fact that we expect the parameters to be similar across treatment groups, and would amount to shrinking the heterogeneous model closer to the homogeneous model. This sort of shrinkage has been shown to work well in related contexts \citep{hahn2017bayesian}, and merits future research.

\section{Bayesian computation}
\label{sec:computation}

Posterior distributions of all the unknown parameters can be easily obtained via standard gibbs-sampling as each parameter is conditionally conjugate, with the exception of $\theta$, which is easy to sample from since it is univariate. A key component driving the ease of sampling is that the Laplace distribution has the following representation as a scale mixture of Gaussians with an exponential mixing weight. 
\begin{align}
	\frac{\lambda}{2} e^{-\lambda |\beta|} = \int_0^{\infty} \frac{1}{\sqrt{2 \pi \tau^2}} e^{- \beta^2 / 2\tau^2} \frac{\lambda^2}{2} e^{-\lambda^2 \tau^2 / 2}. \nonumber
\end{align}

\noindent This makes the prior distribution of $\boldsymbol{\beta}$ multivariate normal with a covariance matrix equal to diag$(\tau_1^2,\dots,\tau_p^2)$. Full details of this mixture as well as posterior implementation can be found in the supplementary materials. The most important parameter of the procedure is $\lambda_0$, which dictates how strongly parameters are shrunk towards zero when they are included in the spike part of the model, i.e $\gamma_j = 0$. Bayesian inference allows for viable alternatives over cross validation, which is commonly used in the penalized likelihood literature. We will examine estimation of $\lambda_0$ using an empirical Bayes procedure, though it is possible to utilize a fully Bayesian specification that places a prior on $\lambda_0$ as discussed in \cite{park2008bayesian}.

\subsection{Selection of $\lambda_0$}
\label{sec:EM}

In many complex settings, such as the current one, empirical Bayes estimators of tuning parameters can not be done analytically. To alleviate this issue \cite{casella2001empirical} proposed a Monte Carlo based approach to finding empirical Bayes estimates of hyperparameter values. The general idea is very similar to the EM algorithm for estimating missing or unknown parameters, however, expectations in the E-step are calculated using draws from a Gibbs Sampler. In our example, we set $\lambda_0^{(0)} = \lambda_0^*$, a starting value of the algorithm. Then for iteration $k$, set 
\begin{align}
	\lambda_0^{(k)} = \sqrt{\frac{2 \left(p - E_{\lambda_0^{(k-1)}} \left(\sum_j \gamma_j \right) \right)}{\sum_j E_{\lambda_0^{(k-1)}}\left(\tau_j^2 1(\gamma_j = 0)\right)}} \nonumber
\end{align}

\noindent where the expectations are approximated with averages from the previous iteration's Gibbs Sampler. Due to Monte Carlo error, this algorithm will not exactly converge, but rather will bounce around the maximum likelihood estimate. The more posterior samples used during each iteration, the less this will occur. Once this has run long enough and the maximum likelihood estimate of $\lambda_0$ is found, inference can proceed by running the same gibbs sampler with the selected $\lambda_0$. A derivation of this quantity can be found in the supplementary materials. 

\subsection{Posterior mode estimation}
The most natural implementation of the above formulation is within the Bayesian paradigm, where we can obtain samples of $\boldsymbol{\gamma}$ directly. This is advantageous as we can examine $p(\boldsymbol{\gamma} \vert \boldsymbol{D})$, which provides an assessment of model uncertainty and can be used to identify the best-fitting models. In some situations, however, MCMC can become burdensome if $p$ is very large. An alternative approach is to formulate model estimation as a penalized likelihood problem, in which we estimate the posterior mode of the model. While in this paradigm we lose some of the aforementioned features of Bayesian inference, estimation can be done in a fraction of the time. Furthermore, the posterior mode of our model will be sparse, i.e many of the regression coefficients will be estimated to be exactly zero allowing us to quickly perform confounder selection in high-dimensions. The details on the penalized likelihood implementation are in the supplementary materials, where we also show that the posterior mode of $\Delta$ from model \ref{eqn:outcome} is consistent at a rate equal to $O \left(\sqrt{\frac{\text{log} p}{n}}\right)$.

\section{Simulation study}
\label{sec:sims}

In this section we compared our proposed approach with several state of the art alternatives for confounding adjustment in the context of $p\geq n$. We consider three data generating mechanisms: 1) homogeneous treatment effect and sparsity; 2) heterogeneous treatment effect and sparsity; and 3) homogeneous treatment effect and non sparsity. Our goal is always to estimate the average treatment effect. Before we detail the data generating mechanisms to be examined, we describe the approaches being compared. 

\begin{enumerate}
	\item Proposed approach using MCMC, where we estimate $\lambda_0$ using the empirical Bayes approach described in Section \ref{sec:EM} and choose $\delta$ as described in Section \ref{sec:delta} (EM-SSL)
    \item EM-SSL for heterogeneous treatment effect
	\item Outcome lasso that includes treatment and covariates, but only places an $l_1$ penalty on the covariates
         \item Re-fit an unpenalized regression model using the covariates identified by the outcome lasso approach above (Post selection lasso)
	\item Double post selection approach of \cite{belloni2013inference} 
	\item Doubly robust lasso approach of \cite{farrell2015robust} 
	\item Approximate residual de-biasing approach of \cite{athey2016approximate} 
\end{enumerate}
The purpose of this simulation study is to assess the performance of our proposed approach compared to competitors when the true outcome model is linear.  In the context of non linear relationships between the covariates and $T$ or $Y$ none of the methods compared here would perform well. 

For some of these estimators, an extension to heterogeneous treatment effects exist, while others implicitly account for treatment effect heterogeneity. With the exception of our estimator, we will always use the version of the estimator that matches the data generating mechanism, e.g the homogeneous version of the estimators will be used for the homogeneous simulation studies. In all simulations the covariates are drawn from a multivariate normal distribution with marginal variances set to 1 and correlation of 0.6 between all covariates. For each scenario, we compare average percent bias, mean squared error, 95\% interval coverage, and the ratio of the average estimated standard errors and the true standard errors. For our approach, the interval coverage is calculated as the percentage of the time our posterior credible interval covers the true parameter, while all other approaches use frequentist confidence intervals. Finally, we present additional simulation results for differing sample sizes and differing confounding strengths in the supplementary materials, though we found the results to be very similar to those seen here. 

\subsection{Homogeneous treatment effects}
\label{sec:simSparse}
We now examine our approach in a high-dimensional setting where $p=500$ and $n=200$, in which there exists strong confounders, weak confounders, and instruments. We simulate the treatment and outcome from the following models:
\begin{align*}
	Y_i &= T_i + \boldsymbol{X_i} \boldsymbol{\beta} + \epsilon_i \\
	logit(p(T_i = 1)) &= \boldsymbol{X_i \psi}
\end{align*}

\noindent where $\epsilon_i \sim \mathcal{N}(0,1)$. The first 8 elements of $\boldsymbol{\beta}$ are $(1,-1,0.3,-0.3, 0, 0, 1, -1)$, while the remaining elements are drawn from a normal distribution with a standard deviation of 0.1. The first 6 elements of $\boldsymbol{\psi}$ are $(1,-1, 1, -1, 1, -1)$ and the remaining values are set to zero. This leads to a treatment prevalence of 50\%. In this setting, covariates 1 and 2 are strong confounders, covariates 3 and 4 are so called ``weak" confounders that are weakly associated with the outcome and strongly associated with treatment, covariates 5 and 6 are instruments, and covariates 7 and 8 are strong predictors of the outcome. The remaining covariates have no association with the treatment and a small to moderate association with the outcome. This situation is not strictly sparse due to the small signals in $\boldsymbol{\beta}$, however, it is approximately sparse in the sense that only a small number of covariates are needed to obtain unbiased estimates of the treatment effect. 

\begin{table}[ht]
\centering
\resizebox{12.5cm}{!}{
\begin{tabular}{lrrrr}
  \hline
type & \% Bias & MSE & 95\% interval coverage & $E(\widehat{SE}(\widehat{\beta_t}))/SD(\widehat{\beta_t})$ \\ 
  \hline
Outcome lasso & 49.1 & 0.34 &  &  \\ 
  Post selection lasso & 27.8 & 0.18 & &  \\ 
  Double post selection & 16.8 & 0.15 & 0.81 & 0.78 \\ 
  Approximate residual de-biasing & 43.2 & 0.28 & 0.73 & 1.03 \\ 
  Doubly robust lasso & 18.7 & 0.26 & 0.72 & 0.64 \\ 
  EM-SSL & 12.9 & 0.10 & 0.93 & 1.01 \\ 
  EM-SSL Heterogeneous & 22.5 & 0.15 & 0.88 & 1.04 \\ 
   \hline
\end{tabular}
}
\caption{Results for estimating the average treatment effect under the simulation scenario of Section \ref{sec:simSparse}}
\label{tab:ACE}
\end{table}

Table \ref{tab:ACE} shows the results of the proposed simulation across 1000 simulated datasets, and we see that the proposed approach performs the best with respect to all metrics. EM-SSL achieves the minimum bias of 12.9\% and the minimum MSE of 0.10. The heterogeneous version of the EM-SSL procedure performs slightly worse in terms of bias and efficiency, which is to be expected given that it splits the sample into the treated and controls and estimates models separately in the two groups. The next best performing estimator in terms of bias and MSE was the double post selection procedure that had a bias of 16.8\% and an MSE of 0.15. The doubly robust lasso had a small bias of 18.7\%, though it was quite variable due to the instability of weights in high-dimensions. In terms of interval estimation we do the best in terms of 95\% interval coverages (93\%) whereas all the other estimators have coverages well below the nominal level (81\%, 73\%, and 72\%). Looking at the ratio of the average estimated to true standard errors, our approach does well (1.01), while most procedures were substantially smaller than 1. The approximate residual de-biasing procedure does well at estimating the standard errors, but is too biased to achieve good interval coverages. Our EM-SSL Heterogeneous procedure also does well at estimating the standard errors, but has an interval coverage of 88\% due to the larger amount of bias relative to the homogeneous version.

\begin{figure}[h]
\centering
 \includegraphics[width = 0.95\linewidth]{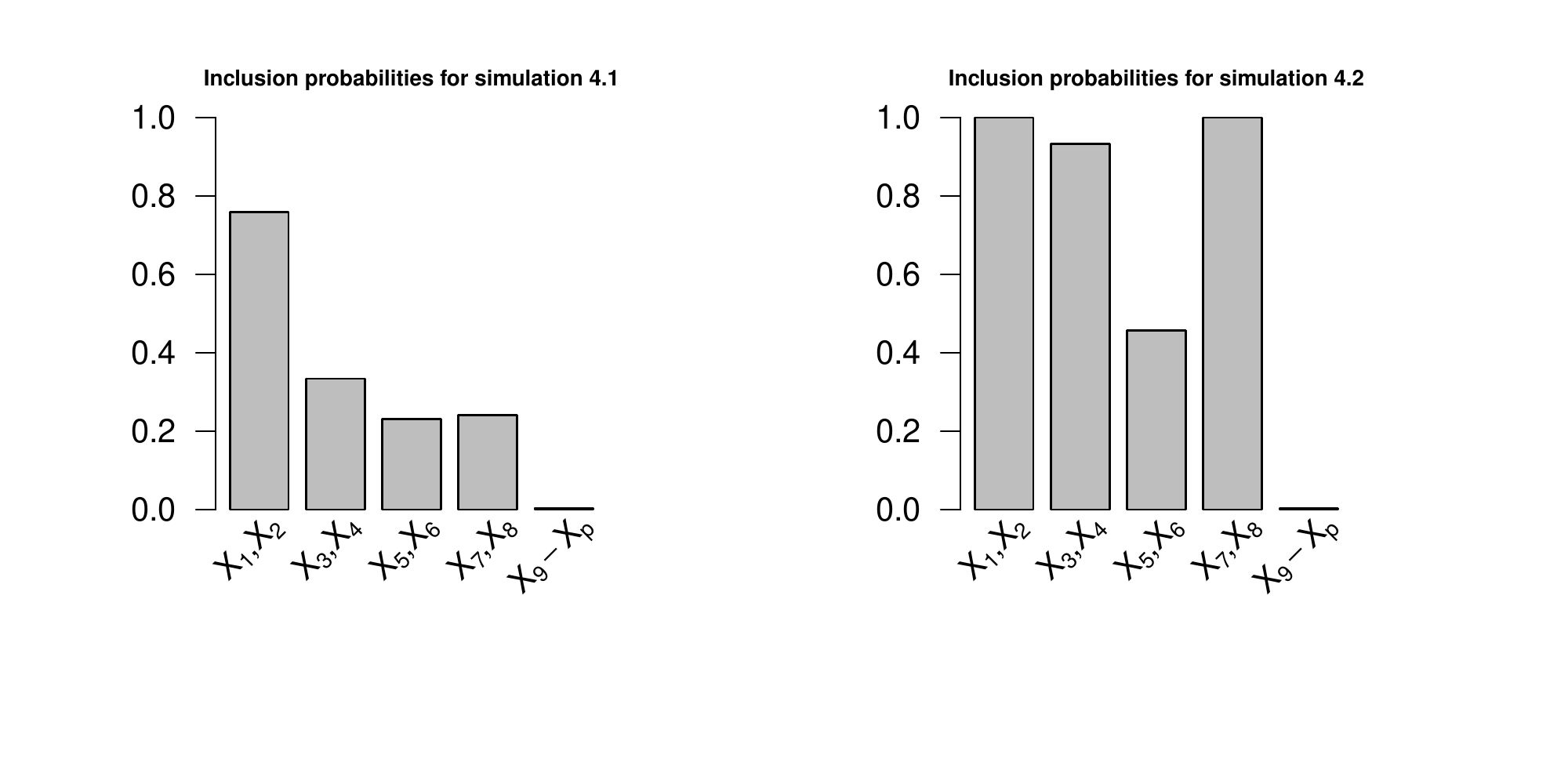}
\caption{Posterior inclusion probabilities from the homogeneous model for simulations in section 4.1 and 4.2.}
\label{fig:probs}
\end{figure}

Figure \ref{fig:probs} shows the posterior inclusion probabilities for the homogeneous EM-SSL model for each of the different types of covariates in the model. We see that the variables strongly associated with both the treatment and outcome ($X_1$ and $X_2$) have the highest value of $P(\gamma_j = 1 \vert \boldsymbol{D})$. Variables $X_3$ and $X_4$ have weak associations with the outcome, but strong relationships with the treatment and they enter into the slab the next highest percentage of the time. Due to our weights, strong instrumental variables ($X_5$ and $X_6$) are in the slab approximately 20\% of the time. Strong predictors of the outcome enter the slab slightly more often than instruments, while the remaining variables almost never enter the slab. These posterior inclusion probabilities highlight why there exists bias in our estimates of the treatment effect. The important confounders are included in the spike component of the prior during some MCMC scans leading to more shrinkage of important components of $\boldsymbol{\beta}$ and biased estimates of the ATE. It is important to note that even when a coefficient is included in the spike, it is not eliminated from the model completely, but rather is more aggressively shrunk to zero. This small amount of bias seems to come with improved efficiency, however, as our estimator has the smallest MSE overall.

\subsection{Heterogeneous treatment effects}
\label{sec:simHetero}

We now simulate data with $n=400$ and $p=800$ and a heterogeneous treatment effect. The data generating models are of the following form:

\begin{align}
	Y_i &= T_i + 0.6 (T_i \times X_{1i}) + 0.4 ( T_i \times X_{3i}) + X_{1i} - X_{2i} + \nonumber\\
	& 0.3 X_{3i} - 0.3 X_{4i} + X_{7i} - X_{8i} + \epsilon_i \nonumber \\
	logit(p(T_i = 1)) &= -1.5 + X_{1i} - X_{2i} + X_{3i} - X_{4i} + X_{5i} - X_{6i} \nonumber
\end{align}

where $\epsilon_i \sim \mathcal{N}(0,1)$. This simulation is similar to the previous section with three changes: 1) covariates 9 through 500 have no association with either treatment or outcome; 2) there is an interaction between the treatment and covariates 1 and 3; and 3) the prevalence of the treatment has been dropped from approximately 50\% to 25\%. We have increased the sample size from $200$ to $400$ since we lowered the prevalence of the treatment, and all methods explored need a sufficient sample size in both the treated and control groups to estimate heterogeneous treatment effects. Table \ref{tab:ACEhetero} shows the results averaged across 1000 simulations. Results are similar to the homogeneous treatment effects setting. The double post selection approach again fares relatively well across all metrics, however, is outperformed by the EM-SSL approach. The approximate residual de-biasing approach has fairly substantial amounts of bias in this setting, which also leads to poor interval coverages. The doubly robust lasso again has a higher MSE due to the instability of inverse propensity weights in high-dimensional settings. Our EM-SSL procedure does not achieve the nominal interval coverage in this setting, though this is due to the bias incurred by assuming a homogeneous treatment effect. Our EM-SSL heterogeneous procedure achieves interval coverages of 95\% and an average estimated standard error that is close to the truth (0.99). Figure \ref{fig:probs} shows the posterior inclusion probabilities for the homogeneous EM-SSL model, and we see that all the important confounders and predictors are included nearly 100\% of the time into the slab component of the prior. This shows that the increased sample size in this simulation, compared with the previous simulation, leads to the improved variable selection. Further, it shows that the bias we see in the EM-SSL estimator is not caused by shrinkage of important parameters, but rather because it assumes homogeneity when the treatment effect is truly heterogeneous.

\begin{table}[ht]
\centering
\resizebox{12.5cm}{!}{
\begin{tabular}{lrrrr}
  \hline
type & \% Bias & MSE & 95\% interval coverage & $E(\widehat{SE}(\widehat{\beta_t}))/SD(\widehat{\beta_t})$ \\ 
  \hline
  Outcome LASSO & 52.0 & 0.30 &  &  \\ 
  Post LASSO & 25.2 & 0.09 &  &  \\ 
  Double post selection & 9.0 & 0.10 & 0.85 & 0.67 \\ 
  Approximate residual de-biasing & 35.0 & 0.16 & 0.45 & 0.93 \\ 
  Doubly robust LASSO & 18.0 & 0.14 & 0.73 & 0.69 \\ 
  EM-SSL & 17.0 & 0.05 & 0.76 & 0.95 \\ 
  EM-SSL Heterogeneous & 5.0 & 0.04 & 0.95 & 0.99 \\    \hline
\end{tabular}
}
\caption{Results for estimating the average treatment effect under the simulation scenario of Section \ref{sec:simHetero}}
\label{tab:ACEhetero}
\end{table}

\subsection{Dense treatment model}
\label{sec:simDense}

Here we simulate data as described in \cite{athey2016approximate} where the treatment model is purposely chosen to be dense. More specifically, first we define 20 clusters, $\{\mathbf{c_1},\dots, \mathbf{c_{20}} \}$ where $\mathbf{c_k} \sim \mathcal{N}(0, I_{p x p})$. Second, we draw $\mathbf{C}_i$ uniformly at random from one of the 20. Third, we draw the covariates from a multivariate normal distribution centered at $\mathbf{C}_i$ with the identity matrix as the covariance. Fourth, we set $T_i = 1$ with probability 0.1 for the first 10 clusters, and $T_i = 1$ with probability 0.9 for the remaining clusters. Finally, we generate data from the outcome model defined as $Y_i = 10 T_i + \boldsymbol{X \beta} + \epsilon_i$, where $\boldsymbol{\beta} \propto (1, \frac{1}{\sqrt{2}}, \dots, \frac{1}{\sqrt{p}})$ and is normalized such that $|| \boldsymbol{\beta}||_2^2 = 18$. Here we will again set $n=200$ and $p=500$. Intuitively, this is a simulation scenario in which the outcome model is approximately sparse, though the treatment model is dense as all of the covariates are associated with the treatment. 
 
Results are summarized in Table 3. Because the data generating mechanism does not assume sparsity, our original EM-SSL procedure performs poorly relative to the post selection lasso approach. We obtain an MSE of 1.07, while also doing very poorly at estimating the standard errors of our approach as the ratio of the average estimated to true standard errors is 0.59. However, under a non sparse setting, if we impose a restriction that only the top $k=10$ variables most associated with the treatment (identified by the magnitude of their coefficients in the treatment lasso model) are prioritized with $w_j = \delta$, then our approach (EM-SSL Restricted) performs the best in terms of MSE (0.59) and interval coverage (93\%). It is also important to note that while we did not show the restricted results in the other simulation scenarios that had sparse treatment models, the restricted approach performed almost identically to the original EM-SSL approach. While there is no principled way of selecting $k$, we have found that other values of $k$, such as $k=20$, perform similarly well.

\begin{table}[ht]
\centering
\resizebox{12.5cm}{!}{
\begin{tabular}{lrrrr}
  \hline
type & \% Bias & MSE & 95\% interval coverage & $E(\widehat{SE}(\widehat{\beta_t}))/SD(\widehat{\beta_t})$ \\ 
  \hline
Outcome LASSO & 0.0 & 0.88 &  &  \\ 
  Post LASSO & 0.0 & 0.76 & &  \\ 
  Double post selection & 0.0 & 1.06 & 0.82 & 0.69 \\ 
  Approximate residual de-biasing & 0.0 & 1.22 & 0.81 & 0.69 \\ 
  Doubly robust lasso & 0.0 & 1.85 & 0.49 & 0.40 \\ 
  EM-SSL & 0.0 & 1.07 & 0.74 & 0.59 \\ 
  EM-SSL Heterogeneous & 0.0 & 1.64 & 0.88 & 0.79 \\ 
  EM-SSL Restricted & 0.0 & 0.59 & 0.93 & 0.93 \\ 
  EM-SSL Restricted Heterogeneous & 0.0 & 1.11 & 0.93 & 0.97 \\  
   \hline
\end{tabular}
}
\caption{Results for estimating the average treatment effect under the simulation scenario of Section \ref{sec:simDense}}
\label{tab:NotSparse}
\end{table}

\subsection{Choosing between homogeneous and heterogeneous models}

An important question is how to decide between the homogeneous and heterogeneous versions of our model in practice. One potential solution to this is to use the WAIC model selection criterion \citep{watanabe2010asymptotic,gelman2014bayesian}, which is a Bayesian analog to traditional model selection tools. We applied WAIC to each of the three simulation scenarios described above to evaluate its effectiveness in choosing the right model. The correct model in Sections \ref{sec:simSparse} and \ref{sec:simDense} is the homogeneous model, and the correct model in Section \ref{sec:simHetero} is the heterogeneous one. In the three simulation scenarios, the WAIC chose the correct model 93\%, 82\%, and 99\% of the time, respectively. This shows that it is possible to automate the decision between the homogeneous and heterogeneous versions of the model, leading to reductions in bias and MSE. These results should be taken with caution, however, as credible intervals from a model chosen using WAIC will not account for the additional uncertainty incurred from the model selection process. This didn't seem to impact our results greatly as our credible interval coverages were 93\%, 92\%, and 94\% for the three simulations explored when using the model chosen by WAIC. 

\section{Analysis of NHANES data}
\label{sec:nhanes}

Recent work \citep{wild2005complementing,patel2010environment,louis2012exposome,patel2012systematic,patel2014studying} has centered on studying the effects of a vast set of exposures on disease. These analyses, termed environmental wide association studies (EWAS), examine environmental factors and aim to improve understanding of the long term effects of different exposures and toxins that humans are invariably exposed to on a daily basis. The National Health and Nutrition Examination Survey (NHANES), is a cross-sectional data source made publicly available by the Centers for Disease Control and Prevention (CDC). The data has also been aggregated and made available by \cite{patel2016database}. The NHANES data is a nationally representative study, in which participants were questioned regarding their health status, with a subset of these patients providing extensive clinical and laboratory tests to provide information on a variety of environmental attributes such as chemical toxicants, pollutants, allergens, bacterial/viral organisms, and nutrients \citep{patel2010environment}.

Our analysis will center on data from the 1999-2000, 2001-2002, 2003-2004, and 2005-2006 surveys. We build on the analysis described in \cite{patel2012systematic}, by applying our proposed methodology to estimating the effects of volatile compounds (VCs) on triglyceride levels in humans. VCs were measured in $n=177$ subjects, and there were $p=127$ covariates. The list of potential confounders consists of other volatile compounds, their interactions, other persistent pesticides measured in the respective subsample, body measurements, demographic, and socioeconomic variables. To evaluate the proposed approach in a setting with $p > n$, we ran an additional analysis, which looked at the effect of VCs on triglycerides in subjects over 40 years old. This led to a sample size of $n=77$ subjects.

In previous work \citep{patel2012systematic} these exposures were evaluated individually without controlling for the remaining pesticides, and only a small subset of pre-selected covariates such as age, BMI, and gender were controlled for. \cite{patel2014studying} wrote that persistent pesticides tend to be highly correlated with one another and that many of the associations found by previous exposome studies \citep{patel2012systematic} could simply be to confounding bias that was unadjusted for due to the small set of confounders used and the fact that other pesticides were not adjusted for. This highlights the need for an analysis that adjusts for all potential confounders, but in our analyses we have $p=127$ covariates with small sample sizes. Therefore, due to the large model space, some confounder selection or shrinkage is required to obtain efficient estimates of exposure effects. 

\subsection{VC analyses}

We now analyze the NHANES data described above. In particular, we will examine the effect of each of 10 volatile compounds on triglyceride levels while controlling for an extensive set of potential confounders. We analyze the effect of each volatile compound using three approaches: 1) an unadjusted model that regresses the outcome on the treatment without confounder adjustment, 2) the homogeneous EM-SSL procedure described above, and 3) the double post selection approach described in \cite{belloni2013inference}. The approximate residual de-biasing and the doubly robust lasso approaches are only applicable to categorical treatments and are therefore left out. We restrict attention to the homogeneous treatment effect application of our approach, because addressing heterogeneity in the setting of continuous treatments would require additional work that is beyond the scope of this paper. 

\begin{figure}[htbp]
\centering
\includegraphics[width=0.9\linewidth]{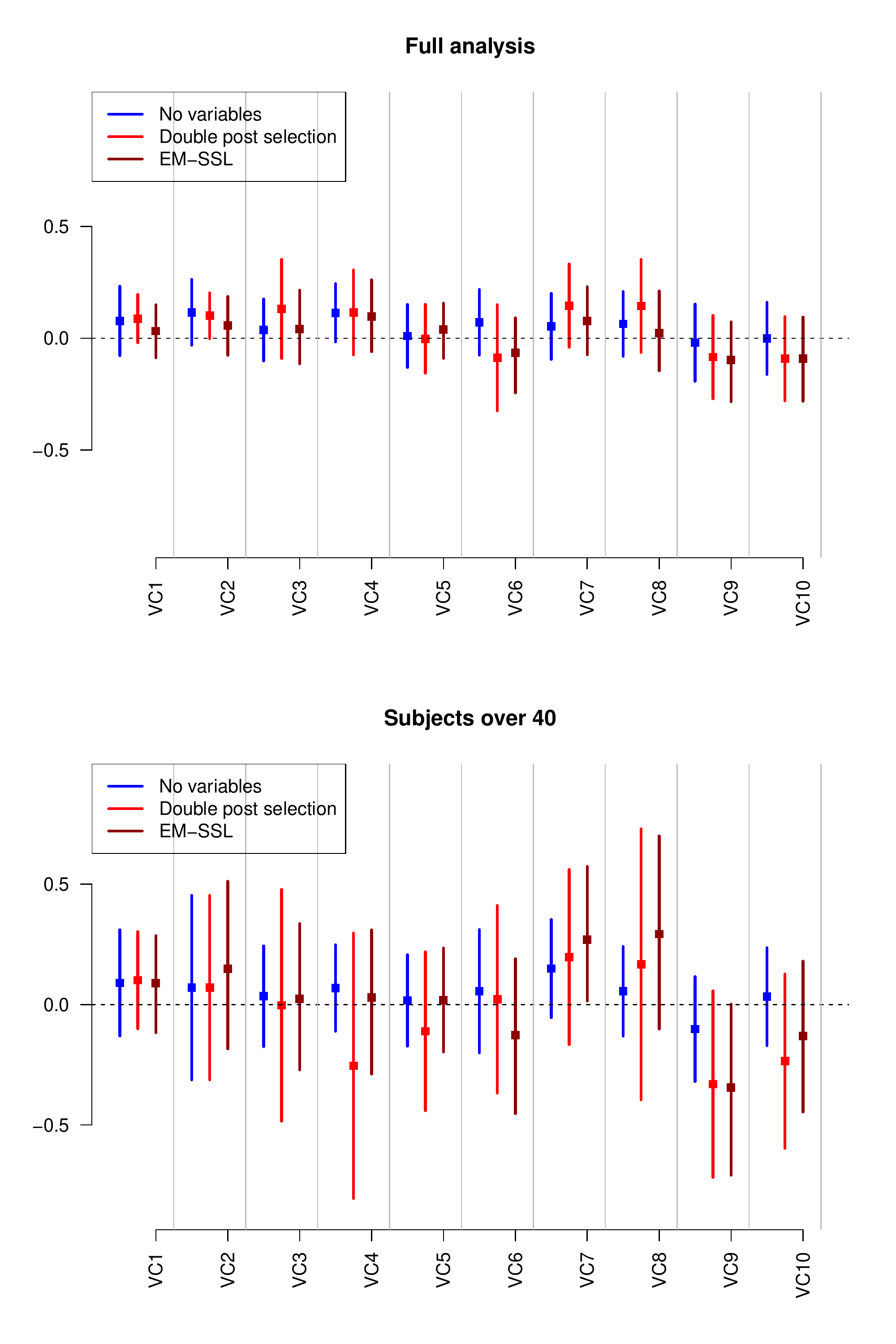}
\caption{Results from analysis of volatile compounds on triglycerides. The upper panel shows the results using the full, $n=177$, sample. The lower panel shows the results for just the $n=77$ subjects who are over 40 years old.}
\label{fig:VCtri}
\end{figure}


Figure \ref{fig:VCtri} shows the point estimates and 95\% confidence intervals (credible interval for EM-SSL approach) from the analysis across the 10 volatile compounds for the three approaches under consideration and each of the two data sets being analyzed. The results are qualitatively very similar across the three approaches for each of the 10 exposures we looked at, in both the full data set and the data set restricting to older subjects. One exception is VC7 in the analysis of older subjects where the EM-SSL estimate has a credible interval that does not contain zero, while the confidence intervals for the other two approaches do. In general, however, the results are very similar in magnitude and direction across the approaches, with the only major difference coming in the widths of the corresponding confidence (and credible) intervals. 

\subsection{Comparison of standard errors across approaches}

\begin{figure}[htbp]
\centering
\includegraphics[width=0.85\linewidth]{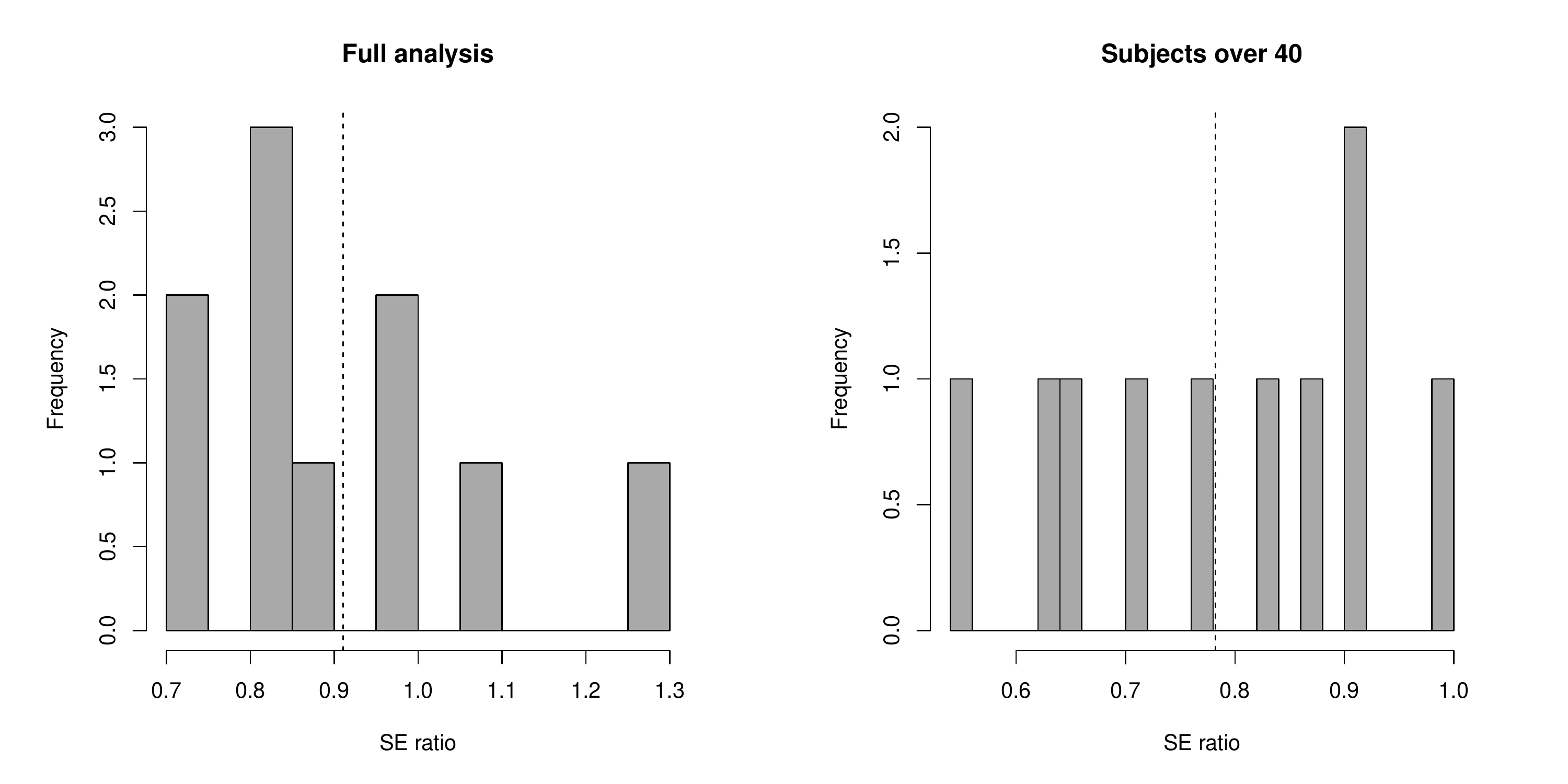}
\caption{The left panel shows a histogram of the ratios of standard errors for the EM-SSL approach and the double post selection approach for the analysis of volatile compounds in the full data. The right panel shows the corresponding histogram for the analysis of subjects over the age of 40. The dashed vertical line is the mean of the ratios that make up the histogram.}
\label{fig:SEratios}
\end{figure}

While there are not drastic differences in point estimates, there are large differences in the widths of the confidence intervals of the two approaches that aim to adjust for confounding. Of interest is the ratio of the standard errors for the EM-SSL procedure and the double post selection approach. These can be seen in Figure \ref{fig:SEratios}. We see that in the full data set the EM-SSL procedure is more efficient overall than the double post selection approach. The majority of the analyses (8/10) had smaller confidence intervals under the EM-SSL procedure with an average confidence interval ratio of 0.9 as indicated by the dashed line in Figure \ref{fig:SEratios}. This means that the EM-SSL procedure on average has a 10\% smaller standard error than the double post selection approach and occasionally has a standard error 30\% smaller. The results are even more striking when we subset the data to the $n=77$ subjects who are over 40 years old. In this case all analyses were more efficient using the EM-SSL procedure, with an average standard error ratio of 0.78, highlighting the ability of our estimator in high-dimensional scenarios. That our estimator is more efficient than the double post selection estimator is not surprising. The goal of the double post selection estimator was to obtain valid inference in high-dimensional scenarios, not to provide the most efficient estimate of the treatment effect. Nonetheless, this analysis highlights an important difference between the approaches in their finite sample performance and how they address instrumental variables. 

\section{Discussion}
\label{sec:discussion}

In this paper we have introduced a novel approach for estimating treatment effects in high-dimensional settings. We introduced a generalization of the spike and slab formulation to allow the prior probability that a parameter for a given covariate is included in the slab component of the prior to depend on the association between each potential confounder and the treatment. We highlighted how this could drastically reduce the shrinkage of important confounders, while still shrinking to zero the coefficients of instrumental and noise variables. Through simulation we showed that our proposed approach has better performance than state of the art approaches under data generating mechanisms that are more or less sparse and also in the context of  heterogeneous treatment effects. By tackling the problem within the Bayesian paradigm we achieve good interval coverage rates even in small samples unlike existing approaches in the literature.  Importantly, we applied the proposed approach to an exposome study and found that our approach gave smaller confidence intervals than existing approaches for confounding adjustment in high dimensional settings.

Our prior is purposely constructed to improve small sample performance of ATE estimation. It shares some commonalities with  doubly robust approaches that aim to model both the treatment and outcome to reduce bias. A crucial difference, however, is that we try to eliminate variables only associated with the treatment while still prioritizing the inclusion of potential confounders in  the outcome model, to minimize both bias and variance in small and finite samples. This differs from existing approaches \citep{belloni2013inference}, which aims at eliminating confounding bias by including all variables associated with either the treatment or outcome. Further, a nice feature of Bayesian approaches is that they account for all the sources of uncertainty in the estimation of the ATE, thus performing better in finite samples than asymptotic approaches. This is because asymptotic approaches often assume that higher-order terms from asymptotic expansions are asymptotically negligible. Such assumptions do not hold in finite samples. Bayesian approaches do not rely on asymptotic expansions nor on the assumption of negligibility detailed above. Statistical uncertainty associated with all the model parameters is indeed accounted for in  the credible intervals for the treatment effect, leading to improved finite sample coverage as seen in Section \ref{sec:sims}. As the sample size increases, the differences between our Bayesian approach and approaches based on asymptotic approximations will diminish.

Our proposed approach has limitations. First, we make the strong assumption of a linear outcome model which allows us to: 1) handle ultra high-dimensional covariate spaces; and 2) borrow information from the treatment model when estimating the causal effects. While similar assumptions of linearity are also used in existing approaches to high-dimensional confounding adjustment \citep{belloni2013inference,farrell2015robust,athey2016approximate,antonelli2016double,shortreed2017outcome}, a topic of future research would be to extend these ideas to nonlinear settings to overcome challenges inherent to model misspecification. Second, we also make the assumption of sparsity of both the treatment and outcome models. While we showed that we can potentially overcome a lack of sparsity in the treatment model by restricting that only a small percentage of the total number of covariates be prioritized in the outcome model, our approach still relies on sparsity of the outcome model to obtain good results in terms of estimation and interval coverages. Third, although we consider scenarios of large $p$, MCMC can become computationally intensive in ultra-high dimensions where the number of covariates is in the tens of thousands. In this setting, alternative approaches such as double post selection, the doubly robust lasso, and approximate residual de-biasing could be used. Lastly, a common criticism of confounder selection is that there is a nonzero probability of excluding a confounder, which leads to bias in the estimation of the causal effect. While excluding a confounder is certainly an issue, we have constructed our prior to avoid this problem as much as possible, by only excluding confounders when this will have very low probability of contributing to bias. In small samples a small amount of bias can be acceptable in high-dimensional scenarios to improve efficiency.

Our prior construction involves building a lasso model as a pre-processing step to build an informative prior. From a Bayesian regression modeling perspective, using the covariates in a regression model to inform the prior distribution of the regression parameters is widely accepted whenever the covariates can be treated as fixed. In standard regression models with covariates $\boldsymbol{X}$, Zellner's g-prior is frequently used, which sets the prior on the variance proportional to $(\boldsymbol{X}^T \boldsymbol{X})^{-1}$. In our approach, the outcome model is treating both $T$ and $\boldsymbol{X}$ as fixed. Therefore constructing a prior from the association between $T$ and $\boldsymbol{X}$ is consistent with standard practice, as long as we don't use the outcome $Y$ to inform the prior. From a causal inference perspective, a prevalent perspective is to separate the design and analysis stages of causal inference \citep{rubin2008objective}. The design phase consists of any steps that occur before analyzing the outcome and can include propensity score modeling, building of matched sets, design of the study, etc. In this perspective, uncertainty in the design phase is typically not accounted for in the analysis phase. Our model follows this perspective, as our prior construction relies on only the propensity score model. Alternative perspectives are now emerging \citep{liao2018uncertainty}, and extensions of our approach would be worthy of consideration for future work.

There are a number of extensions of the proposed ideas that merit further research. In the current manuscript we restricted attention to the case where $w_j = \delta$ for all variables associated with the treatment. We can relax this assumption to let $w_j$ vary for each covariate $j$ associated with the treatment, potentially improving on the current approach, though future research would be required to find an optimal strategy. The ideas in this paper could also be used to improve finite sample estimation for doubly robust estimators. Doubly robust estimators typically combine an outcome regression and a treatment model, and our ideas could be used to improve the outcome model in this estimator. This, coupled with improved estimation of the treatment model using similar ideas as done in \cite{shortreed2017outcome}, could lead to improved doubly robust estimators. Finally, the idea to borrow information from the treatment model to guide the amount of shrinkage in the outcome model can be extended to other high-dimensional priors beyond the spike and slab one seen here. A number of priors are used in high-dimensional Bayesian modeling, and this idea can potentially be extended to many of them.

\section*{Acknowledgements}
The authors are grateful for Chirag Patel and his advice regarding the NHANES data analysis. Funding for this work was provided by National Institutes of Health (ES000002, ES024332, ES007142, ES026217 P01CA134294, R01GM111339, R35CA197449, P50MD010428)

\section*{Software}

An R package implementing the methodology proposed here can be found at \\ github.com/jantonelli111/HDconfounding

\appendix

\section{Posterior computation}

The laplace distribution has the following representation as a scale mixture of Gaussians with an exponential mixing weight. 

\begin{align}
	\frac{\lambda}{2} e^{-\lambda |\beta|} = \int_0^{\infty} \frac{1}{\sqrt{2 \pi \tau^2}} e^{- \beta^2 / 2\tau^2} \frac{\lambda^2}{2} e^{-\lambda^2 \tau^2 / 2}.
\end{align}

\noindent This makes the conditional distribution of $\boldsymbol{\beta}$ multivariate normal. To simplify notation we will also define $\boldsymbol{X}^* = [\boldsymbol{1}', \boldsymbol{T}', \boldsymbol{X}]$ and $\boldsymbol{\beta}^* = [\beta_0, \beta_t, \boldsymbol{\beta}]$. Conditional on a value of $\lambda_0$ the algorithm iterates through the following steps:

\begin{enumerate}
	\item Sample $\boldsymbol{\beta}^*$ from the full conditional:
    \begin{align*}
    	\boldsymbol{\beta}^* \vert \bullet \sim \text{Normal} \left( \left({\boldsymbol{X}^*}^T \boldsymbol{X}^* + \boldsymbol{D_\tau}^{-1} \right)^{-1} {\boldsymbol{X}^*}^T \boldsymbol{Y}, \sigma^2 \left({\boldsymbol{X}^*}^T \boldsymbol{X}^* + \boldsymbol{D_\tau}^{-1} \right)^{-1}\right)
    \end{align*}
    where $\boldsymbol{D_\tau}$ is a diagonal matrix with diagonal $(K, K, \tau_1^2, \dots, \tau_p^2)$. To speed up computation time this can be broken up into block updates instead of all $p$ variables at once.
    
    \item For $j=1,\dots,p$ sample $\gamma_j$ from a bernoulli distribution with probability
    \begin{align*}
    	\frac{\frac{\lambda_1}{\sigma} e^{\frac{-\lambda_1 |\beta_j|}{\sigma}} \theta^{w_j}}{\frac{\lambda_1}{\sigma} e^{\frac{-\lambda_1 |\beta_j|}{\sigma}} \theta^{w_j} + \frac{\lambda_0}{\sigma} e^{\frac{-\lambda_0 |\beta_j|}{\sigma}} (1 - \theta^{w_j})}
     \end{align*}
     \item Use a random-walk Metropolis Hastings algorithm, rejection sampling, or any other approach to sampling from a univariate distribution without a closed form, to sample from the full conditional of $\theta$
     \begin{align*}
        	p(\theta \vert \bullet) \propto \theta^{a + \sum_{j=1}^p w_j \gamma_j} (1 - \theta)^b \prod_{j=1}^p (1 - \theta^{w_j})^{(1 - \gamma_j)}
     \end{align*}
     \item Letting $\eta_j^2 = 1/\tau_j^2$, sample from the full conditional:
     \begin{align*}
     	\eta_j^2 \vert \bullet \sim \text{InvGauss} \left(\mu' = \sqrt{\frac{\lambda' \sigma^2}{\beta_j^2}}, \lambda' = (\lambda_1 \gamma_j + \lambda_0(1 - \gamma_j))^2 \right)
     \end{align*}
     where the inverse-Gaussian density is given by
     \begin{align*}
     	f(x) = \sqrt{\frac{\lambda'}{2 \pi}} x^{-3/2} exp \left\{\frac{\lambda'(x - \mu')^2}{2 {\mu'}^2 x} \right\}
     \end{align*}
     \item Sample $\sigma^2$ from the full conditional
     \begin{align*}
     	\sigma^2 \vert \bullet \sim \text{InvGamma} \left(c + \frac{n}{2} + \frac{p}{2}, d + \frac{|| \boldsymbol{Y} - \boldsymbol{X^* \beta^*}||}{2} + \frac{\boldsymbol{\beta}^T \boldsymbol{D}_{\tau, \beta}^{-1} \boldsymbol{\beta} }{2} \right)
     \end{align*}
     Where $\boldsymbol{D}_{\tau, \beta}$ is a diagonal matrix with diagonal $(\tau_1^2, \dots, \tau_p^2)$
\end{enumerate}

\section{Finding empirical Bayes estimate of $\lambda_0$}

As seen in \cite{casella2001empirical} the Monte Carlo EM algorithm treats the parameters as missing data and iterates between finding the ``complete data'' log likelihood where the parameter values are filled in with their expectations from the gibbs sampler, and maximizing this expression as a function of the hyperparameter values. After removing the terms not involving $\lambda_0$ the ``complete data'' log likelihood can be written as

\begin{align}
	(p - \sum_{j=1}^p \gamma_j) ln(\lambda_0^2) - \frac{\lambda_0^2}{2} \sum_{j=1^p} \tau_j^2 1(\gamma_j = 1)
\end{align}

And now, adopting the same notation that is typically used with the EM algorithm and was seen in \cite{park2008bayesian} we can take the expectation of this expression, where expectations are taken as the averages over the previous iterations posterior samples

\begin{align}
	Q(\lambda_0 | \lambda_0^{(k-1)}) = \left(p - \sum_{j=1}^p E_{\lambda_0^{(k-1)}}(\gamma_j)\right) ln(\lambda_0^2) - \frac{\lambda_0^2}{2} \sum_{j=1}^p E_{\lambda_0^{(k-1)}} \left( \tau_j^2 1(\gamma_j = 1) \right)
\end{align}

And now we can find the maximum of this expression with respect to $\lambda_0$ and we find that the maximum occurs at 

\begin{align}
	\lambda_0^{(k)} = \sqrt{\frac{2 \left(p - E_{\lambda_0^{(k-1)}} \left(\sum_j \gamma_j \right) \right)}{\sum_j E_{\lambda_0^{(k-1)}}\left(\tau_j^2 1(\gamma_j = 0)\right)}}
\end{align}

\section{Posterior mode estimation}

In the manuscript we briefly described the details involved with estimating the posterior mode of our model. Here we provide a more detailed description of the coordinate ascent algorithm. These ideas were first introduced in \cite{rovckova2016spike}, and we extend them here to differing weights, $w_j$. We will defer many of the implementation details to \cite{rovckova2016spike}, though we will highlight many of them here to illustrate how the weights, $w_j$ can reduce shrinkage for potential confounders. \\
\\
\textbf{Assuming $\theta$ is fixed} \\
\\
Adopting similar notation to \cite{rovckova2016spike} we can formulate the problem in the standard penalized regression framework

\begin{align*}
	(\widehat{\beta_0}, \widehat{\beta_t}, \boldsymbol{\widehat{\beta}}) = \underset{(\beta_0, \beta_t, \boldsymbol{\beta}) \in \mathbb{R}^{p+2}}{arg max} \left \{ -\frac{1}{2} || \boldsymbol{Y} - \boldsymbol{X\beta} - \beta_t \boldsymbol{T} - \beta_0 ||^2 + pen(\boldsymbol{\beta} \vert \theta) \right \}
\end{align*}

\noindent Notice that in the penalty we have conditioned on a value of $\theta$. To illustrate the results and the implementation of the approach we will first illustrate them for a fixed value of $\theta$. Obviously, we want to let the data inform $\theta$, which will then provide us with multiplicity control \cite{scott2010bayes}. As shown in \cite{rovckova2016spike} the extension to random values of $\theta$ is straightforward and the same results will apply with a minor adjustment. The penalty can be written as

\begin{align}
	pen(\boldsymbol{\beta} \vert \theta) &=  \sum_{j=1}^p \rho(\beta_j \vert \theta)\\
	&= \sum_{j=1}^p -\lambda_1 |\beta_j| + log \left(  \frac{p_{\theta}^*(0)}{p_{\theta}^*(\beta_j)} \right)
\end{align}

\noindent where

\begin{align}
	p_{\theta}^*(\beta_j) = P(\gamma_j = 1 \vert \beta_j, \theta) = \frac{\theta^{w_j} \psi_1(\beta_j)}{\theta^{w_j} \psi_1(\beta_j) + (1 - \theta^{w_j}) \psi_0(\beta_j)}
\end{align}

\noindent An important feature of the penalty can be seen by looking at it's derivative, which gives insight into the level of shrinkage induced for $\beta_j$. 

\begin{align}
	\frac{\partial pen(\boldsymbol{\beta} \vert \theta)}{\partial |\beta_j|} \equiv - \lambda_{\theta}^*(\beta_j),
\end{align}

\noindent where

\begin{align}
	\lambda_{\theta}^*(\beta_j) = \lambda_1 p_{\theta}^*(\beta_j) + \lambda_0 (1 - p_{\theta}^*(\beta_j)).
\end{align}

\noindent The intuition for $\lambda_{\theta}^*(\beta_j)$ is the larger it gets, the more shrinkage that is induced for $\beta_j$. If $p_{\theta}^*(\beta_j)$ is large, then it is likely that covariate $j$ is important and it will get shrunk by a smaller amount. These ideas can be seen by looking at the eventual estimation strategy for estimating $\widehat{\boldsymbol{\beta}}$, which we detail now. We can define

\begin{align}
	\Delta_j \equiv \underset{t>0}{inf} \left\{ nt/2 - \rho(t \vert \theta)/t \right\},
\end{align}

\noindent where there is a subscript $j$, because $\rho(t \vert \theta)$ depends on the weight, $w_j$. Then the global mode, $\widehat{\boldsymbol{\beta}}$ satisfies the following
\begin{align}
\widehat{\beta_j} = \begin{cases}
				0 \ \ \ \ \ \ \ \ \ \ \ \ \ \ \ \ \ \ \ \ \ \ \ \ \ \ \ \ \ \ \ \ \ \text{ if } |z_j| \leq \Delta_j \\
				\frac{1}{n} \left(|z_j| - \lambda_{\theta}^*(\widehat{\beta_j}) \right)_+ \text{sign}(z_j) \text{ if } |z_j| > \Delta_j
			     \end{cases} \label{eqn:ns_solution_app}
\end{align}

\noindent where $z_j = \boldsymbol{X_j}'(\boldsymbol{Y} - \sum_{l \neq j} \boldsymbol{X_l}\widehat{\beta_l} - \boldsymbol{T} \widehat{\beta_t} - \widehat{\beta_0})$. This highlights a couple very important features both for implementation of the approach as well as understanding the role of the prior distribution in penalizing covariates when the goal is confounder selection. This is very useful for implementation, as we can adopt a coordinate ascent algorithm that iterates through each of the covariates and updates them according to (\ref{eqn:ns_solution_app}). We also now see the role that $p_{\theta}^*(\beta_j)$, and therefore $w_j$ plays in both the shrinkage of covariates and the decision to force them to zero or not. As $w_j$ gets smaller, $p_{\theta}^*(\beta_j)$ gets larger and then $\Delta_j$ gets smaller making it less likely that we estimate $\beta_j$ to be zero. For those coefficients that are not thresholded to zero, there still exists soft thresholding or shrinkage and this is dictated by $\lambda_{\theta}^*(\beta_j)$. The smaller we set $w_j$, the smaller  $\lambda_{\theta}^*(\beta_j)$ becomes leading to less shrinkage. This shows that if we can set $w_j$ to be smaller for potential confounders then we have both increased their probability of making it into the final model (i.e $\widehat{\beta_j} \neq 0$), and we have reduced the shrinkage affecting potential confounders. \\
\\
\textbf{Assuming $\theta$ is random} \\
\\
Now that we have illustrated the approach for a fixed value of $\theta$, we can extend it to random $\theta$, by letting the data inform the level of sparsity. \cite{rovckova2016spike} showed a crucial result when attempting to implement a random value for $\theta$. Many of the previous results were allowed because the penalty on $\boldsymbol{\beta}$ was separable, i.e the penalties for each $\beta_j$ were independent. Now they are not independent as the penalty marginalizes over $\theta$, which is estimated from all of the regression coefficients and the penalties are no longer marginally independent. \cite{rovckova2016spike} showed, however, that all of the previous results hold in the situation where $\theta$ is estimated if we replace $\theta$ in all the expressions with $\theta_j = E(\theta \vert \widehat{\boldsymbol{\beta}}_{\backslash j})$ where $\widehat{\boldsymbol{\beta}}_{\backslash j}$ is the vector of estimated regression coefficients with the $j^{th}$ element removed. When $p$ is large then $E(\theta \vert \widehat{\boldsymbol{\beta}}_{\backslash j}) \approx E(\theta \vert \widehat{\boldsymbol{\beta}})$, and therefore we can let the value of $\theta$ remain constant for each $j$. This conditional expectation is given by

\begin{align}
	E(\theta \vert \widehat{\boldsymbol{\beta}}) = \frac{\int_0^1 \theta^a (1 - \theta)^{b-1} \prod_{j=1}^p \left\{ \theta^{w_j} \lambda_1 e^{- | \widehat{\beta_j} | (\lambda_1)} + (1 - \theta^{w_j}) \lambda_0 e^{- | \widehat{\beta_j} | (\lambda_0)} \right\} d\theta}{\int_0^1 \theta^{a-1} (1 - \theta)^{b-1} \prod_{j=1}^p \left\{ \theta^{w_j} \lambda_1 e^{- | \widehat{\beta_j} | (\lambda_1)} + (1 - \theta^{w_j}) \lambda_0 e^{- | \widehat{\beta_j} | (\lambda_0)} \right\} d\theta}. \label{eqn:Etheta_app}
\end{align}		  

\noindent This expression does not have a closed form, but is fairly easy to calculate numerically. Because we can simply take all the implementation steps from above and replace $\theta$ with $E(\theta \vert \widehat{\boldsymbol{\beta}})$, we can update our coordinate ascent algorithm accordingly. Now after each time we update $\beta_j$ for each $j$, we must also update $\theta$ according to (\ref{eqn:Etheta_app}). This can be computationally expensive, particularly in very high-dimensional settings. In reality $E(\theta \vert \widehat{\boldsymbol{\beta}})$, will not change much after each updated $\beta_j$, especially in sparse settings when most of the parameters are not included in the model, i.e $\beta_j = 0$. In practice, one can update $\theta$ in the coordinate ascent algorithm after every $m^{th}$ covariate is updated, where $m$ is some large value that is also smaller than $p$. \\
\\
\textbf{Selection of $\lambda_0$} \\
\\
As with fully Bayesian inference, our coordinate ascent algorithm relies on a well chosen value of $\lambda_0$, and there are two distinct solutions within this paradigm. The first of which, cross validation, is commonly used for penalized likelihood procedures to find tuning parameter values. Cross validation is used to find a value of the tuning parameter that aims to optimize the model's predictive ability. We are interested in confounding adjustment, not prediction, though cross validation still might ultimately be useful for finding an appropriate $\lambda_0$ estimate. Another approach, as described in \cite{rovckova2016spike}, involves estimating the model for an increasing sequence of $\lambda_0$ values and then finding the value of $\lambda_0$ at which the estimates stabilize. We can evaluate $\widehat{\boldsymbol{\beta}}$ for $\lambda_0 \in \left\{\lambda_0^{(1)}, \lambda_0^{(2)}, \dots \right\}$, which is a sequence of penalties that more aggressively penalize coefficients. As shown in \cite{rovckova2016spike} the estimates $\widehat{\boldsymbol{\beta}}$ eventually converge as we increase $\lambda_0$ and we can take the estimates of $\boldsymbol{\beta}$ at this stabilization point as the final posterior mode estimate.

\section{Theoretical considerations}
\label{sec:theory}

\cite{rovckova2016spike} provide a number of asymptotic results justifying the use of the spike and slab lasso prior in high dimensional settings. In particular, they derived asymptotic rates of convergence for $\boldsymbol{\beta}$, such as the fact that $|X(\boldsymbol{\widehat{\beta}} - \boldsymbol{\beta})|_2 < C_1 \sqrt{q(1 + C_2) \text{log } p}$, where $C_1, C_2$ are positive constants and $q = ||\boldsymbol{\beta}||_0$. It is a straightforward application of their proofs to obtain identical results under our prior, however, our interest is in estimating $\beta_t$, not in estimating $\boldsymbol{\beta}$. One can use Equation \ref{eqn:ns_solution_app} to see that
\begin{align}
	\widehat{\beta}_t = \beta_t  + \frac{1}{n} \left( \boldsymbol{T}'(\boldsymbol{X\beta} - \boldsymbol{X\widehat{\beta}}) + \boldsymbol{T'\epsilon}) \right)
\end{align}

\noindent We can then apply the triangle inequalty and H\"{o}lder's inequality to see that
\begin{align}
	|\widehat{\beta}_t - \beta_t|_1 \leq &  \left|\frac{1}{n} \boldsymbol{T}'(\boldsymbol{X\beta} - \boldsymbol{X\widehat{\beta}})\right|_1 + \frac{1}{n}\left| \boldsymbol{T'\epsilon} \right|_1 \\
    \leq & \frac{1}{n} \left| \boldsymbol{T} \right|_2 \left|(\boldsymbol{X\beta} - \boldsymbol{X\widehat{\beta}})\right|_2 + \frac{1}{n}\left| \boldsymbol{T'\epsilon} \right|_1 
    = O \left(\sqrt{\frac{\text{log} p}{n}}\right),
\end{align}
where in the last step we used that $\frac{1}{n}\left| \boldsymbol{T'\epsilon} \right|_1 \rightarrow 0$ and we made the assumption that $T$ has bounded second moments, which is very reasonable, particularly when $T$ is binary where it is certain to hold. This shows that our estimate of the treatment effect converges, but at a slower rate than $n^{1/2}$ as we see the usual log $p$ penalty found in high-dimensional models. Interestingly, \cite{belloni2013inference,farrell2015robust, belloni2017program,athey2016approximate} showed that $n^{1/2}-$consistent estimation can be achieved when estimating treatment effects in high dimensional scenarios. Our approach is mainly targeted at improving finite sample properties instead of asymptotic rates.

\section{Simulations with different sample sizes and confounding strengths}

Here we present simulation results that build on those found in the main manuscript. We will present three different scenarios here: 1) A scenario that is the same as the one of Section 4.1, but with $n=400$ and $p=800$, 2) A scenario that is the same as the simulation in Section 4.3, but with $n=400$ and $p=800$, and 3) A scenario that is the same as the simulation from Section 4.3 where the confounding strength has been increased by setting $|| \boldsymbol{\beta} ||_2 = 18$. Below we present the corresponding tables for each of these scenarios. \\
\\
\newpage
\textbf{Additional scenario 1:} \\
\\
\begin{table}[ht]
\centering
\resizebox{12.5cm}{!}{
\begin{tabular}{lrrrr}
  \hline
type & \% Bias & MSE & 95\% interval coverage & $E(\widehat{SE}(\widehat{\beta_t}))/SD(\widehat{\beta_t})$ \\ 
  \hline
Outcome LASSO & 39.6 & 0.21 &  &  \\ 
  Post LASSO & 24.2 & 0.11 &  &  \\ 
  Double post selection & 14.4 & 0.09 & 0.76 & 0.73 \\ 
Approximate residual de-biasing & 30.1 & 0.14 & 0.76 & 1.04 \\
 Doubly robust lasso & 14.1 & 0.17 & 0.71 & 0.61 \\      EM-SSL & 6.9 & 0.05 & 0.93 & 0.95 \\ 
  EM-SSL Heterogeneous & 9.5 & 0.06 & 0.93 & 1.06 \\ 
  EM-SSL Restricted & 6.4 & 0.05 & 0.93 & 0.95 \\ 
  EM-SSL Restricted Heterogeneous & 9.4 & 0.06 & 0.94 & 1.05 \\   
   \hline
\end{tabular}
}
\caption{Results for estimating the average treatment effect under the first additional simulation scenario.}
\end{table}

\textbf{Additional scenario 2:} \\
\\
\begin{table}[ht]
\centering
\resizebox{12.5cm}{!}{
\begin{tabular}{lrrrr}
  \hline
type & \% Bias & MSE & 95\% interval coverage & $E(\widehat{SE}(\widehat{\beta_t}))/SD(\widehat{\beta_t})$ \\ 
  \hline
Outcome LASSO & 0.0 & 0.36 &  &  \\ 
  Post LASSO & 0.0 & 0.40 &  &  \\ 
  Double post selection & 0.0 & 0.38 & 0.80 & 0.67 \\ 
  Approximate residual de-biasing & 0.0 & 0.47 & 0.85 & 0.79 \\ 
  Doubly robust lasso & 0.0 & 1.42 & 0.67 & 0.50 \\ 
  EM-SSL & 0.0 & 0.34 & 0.79 & 0.67 \\ 
  EM-SSL Heterogeneous & 0.0 & 0.51 & 0.93 & 0.96 \\ 
  EM-SSL Restricted & 0.0 & 0.21 & 0.94 & 1.00 \\ 
  EM-SSL Restricted Heterogeneous & 0.0 & 0.42 & 0.97 & 1.03 \\ 
   \hline
\end{tabular}
}
\caption{Results for estimating the average treatment effect under the second additional simulation scenario.}
\end{table}

\textbf{Additional scenario 3:} \\
\\
\begin{table}[ht]
\centering
\resizebox{12.5cm}{!}{
\begin{tabular}{lrrrr}
  \hline
type & \% Bias & MSE & 95\% interval coverage & $E(\widehat{SE}(\widehat{\beta_t}))/SD(\widehat{\beta_t})$ \\ 
  \hline
Outcome LASSO & 0.0 & 14.28 &  &  \\ 
  Post LASSO & 0.0 & 12.31 & &  \\   
Double post selection & 0.0 & 15.08 & 0.82 & 0.72 \\   Approximate residual de-biasing & 0.0 & 21.44 & 0.80 & 0.66 \\ 
Doubly robust lasso & 0.0 & 13.75 & 0.43 & 0.37 \\     EM-SSL & 0.0 & 12.25 & 0.82 & 0.71 \\ 
  EM-SSL Heterogeneous & 0.0 & 24.46 & 0.89 & 0.84 \\ 
  EM-SSL Restricted & 0.0 & 8.83 & 0.93 & 0.96 \\ 
  EM-SSL Restricted Heterogeneous & 0.0 & 18.76 & 0.95 & 0.98 \\ 
   \hline
\end{tabular}
}
\caption{Results for estimating the average treatment effect under the third additional simulation scenario.}
\end{table}

\section{Simulations for sensitivity to hyperparameters}

First, we will assess the sensitivity to the choice of $\lambda_1$ by running the same analysis as in Section 4.1 of the manuscript and varying $\lambda_1 \in \{ 0.05, 0.25, 0.5\}$. Below are the tables for the three different values of $\lambda_1$ and we see that the results are highly similar to those in the manuscript showing the robustness to the choice of this parameter. \\

$\boldsymbol{\lambda_1 = 0.05}$
\begin{table}[H]
\centering
\resizebox{!}{1.17cm} {
\begin{tabular}{lrrrr}
  \hline
type & \% Bias & MSE & 95\% interval coverage & $E(\widehat{SE}(\widehat{\beta_t}))/SD(\widehat{\beta_t})$ \\ 
  \hline
EM-SSL & 15.00 & 0.11 & 0.92 & 1.00 \\ 
  EM-SSL Heterogeneous & 22.00 & 0.16 & 0.90 & 1.03 \\ 
  EM-SSL Restricted & 15.00 & 0.11 & 0.91 & 1.00 \\ 
  EM-SSL Restricted Heterogeneous & 24.00 & 0.16 & 0.88 & 1.03 \\ 
   &  &  &  &  \\ 
   \hline
\end{tabular}
}
\end{table}

$\boldsymbol{\lambda_1 = 0.25}$
\begin{table}[ht]
\centering
\resizebox{!}{1.17cm} {
\begin{tabular}{lrrrr}
  \hline
type & \% Bias & MSE & 95\% interval coverage & $E(\widehat{SE}(\widehat{\beta_t}))/SD(\widehat{\beta_t})$ \\ 
  \hline
EM-SSL & 14.00 & 0.10 & 0.93 & 1.01 \\ 
  EM-SSL Heterogeneous & 24.00 & 0.16 & 0.90 & 1.04 \\ 
  EM-SSL Restricted & 14.00 & 0.10 & 0.92 & 1.01 \\ 
  EM-SSL Restricted Heterogeneous & 24.00 & 0.16 & 0.89 & 1.05 \\ 
   &  &  &  &  \\ 
   \hline
\end{tabular}
}
\end{table}

$\boldsymbol{\lambda_1 = 0.5}$
\begin{table}[ht]
\centering
\resizebox{!}{1.17cm} {
\begin{tabular}{lrrrr}
  \hline
type & \% Bias & MSE & 95\% interval coverage & $E(\widehat{SE}(\widehat{\beta_t}))/SD(\widehat{\beta_t})$ \\ 
  \hline
EM-SSL & 14.00 & 0.10 & 0.93 & 1.02 \\ 
  EM-SSL Heterogeneous & 25.00 & 0.16 & 0.89 & 1.07 \\ 
  EM-SSL Restricted & 15.00 & 0.10 & 0.92 & 1.01 \\ 
  EM-SSL Restricted Heterogeneous & 26.00 & 0.16 & 0.88 & 1.06 \\ 
   &  &  &  &  \\ 
   \hline
\end{tabular}
}
\end{table}

Now we also want to assess the sensitivity to the choice of $a$ and $b$ which dictate the prior distribution for $\theta$. Again we will use the same analysis as in Section 4.1 of the manuscript except we will vary the values of $a$ and $b$. In the manuscript we set $a=1$ and $b=0.1p$. Here we will try two different scenarios: $a=1$ and $b=1$, as well as $a=1$ and $b=p$. The first of these scenarios represents a situation where the prior probability of entering into the slab is approximately 0.5, and therefore much less sparse. The second situation is a more stringent prior that more strongly imposes sparsity in the model as it is less likely a priori to enter the slab. We see in the results below that our approach seems to be robust to the choice of these parameters as we obtain similar MSE and interval coverage values on either end of the spectrum of prior sparsity. \\

\textbf{a = 1, b = p}
\begin{table}[H]
\centering
\begin{tabular}{lrrrr}
  \hline
type & \% Bias & MSE & 95\% interval coverage & $E(\widehat{SE}(\widehat{\beta_t}))/SD(\widehat{\beta_t})$ \\ 
  \hline
EM-SSL & 15.00 & 0.11 & 0.92 & 1.00 \\ 
  EM-SSL Heterogeneous & 22.00 & 0.16 & 0.90 & 1.04 \\ 
  EM-SSL Restricted & 15.00 & 0.11 & 0.91 & 1.00 \\ 
  EM-SSL Restricted Heterogeneous & 24.00 & 0.16 & 0.88 & 1.03 \\ 
   &  &  &  &  \\ 
   \hline
\end{tabular}
\end{table}

\textbf{a = 1, b = 1}
\begin{table}[H]
\centering
\begin{tabular}{lrrrr}
  \hline
type & \% Bias & MSE & 95\% interval coverage & $E(\widehat{SE}(\widehat{\beta_t}))/SD(\widehat{\beta_t})$ \\ 
  \hline
EM-SSL & 14.00 & 0.10 & 0.92 & 1.00 \\ 
  EM-SSL Heterogeneous & 22.00 & 0.16 & 0.90 & 1.03 \\ 
  EM-SSL Restricted & 15.00 & 0.10 & 0.92 & 1.00 \\ 
  EM-SSL Restricted Heterogeneous & 24.00 & 0.16 & 0.89 & 1.04 \\ 
   &  &  &  &  \\ 
   \hline
\end{tabular}
\end{table}

\bibliographystyle{authordate1}
\bibliography{SSLconfounding}

\label{lastpage}

\end{document}